# PadSteg: Introducing Inter-Protocol Steganography

Bartosz Jankowski, Wojciech Mazurczyk, Krzysztof Szczypiorski

*Abstract* — **Hiding information in network traffic may lead to leakage of confidential information. In this paper we introduce a new steganographic system: the *PadSteg* (Padding Steganography). To authors' best knowledge it is the first information hiding solution which represents *inter-protocol steganography* i.e. usage of relation between two or more protocols from the TCP/IP stack to enable secret communication. *PadSteg* utilizes ARP and TCP protocols together with an *Etherleak* vulnerability (improper Ethernet frame padding) to facilitate secret communication for hidden groups in LANs (Local Area Networks). Basing on real network traces we confirm that *PadSteg* is feasible in today's networks and we estimate what steganographic bandwidth is achievable while limiting the chance of disclosure. We also point at possible countermeasures against *PadSteg*.**

## I. INTRODUCTION

Network steganography is currently seen as a rising threat to network security. Contrary to typical steganographic methods which utilize digital media (pictures, audio and video files) as a cover for hidden data (steganogram), network steganography utilizes communication protocols' control elements and their basic intrinsic functionality. As a result, such methods may be harder to detect and eliminate.

In order to minimize the potential threat to public security, identification of such methods is important as is the development of effective detection (steganalysis) methods. This requires both an in-depth understanding of the functionality of network protocols and the ways in which it can be used for steganography. Many methods had been proposed and analyzed so far – for the detailed review see Zander et al. [2] or Petitcolas et al. [3].

Typical network steganography method uses modification of a single network protocol. The classification of so such methods was introduced by Mazurczyk et al. in [15]. The protocol modification may be applied to the PDU (Protocol Data Unit) [1], [4], [5], time relations between exchanged PDUs [6], or both [14] (hybrid methods). This kind of network steganography can be called *intra-protocol* steganography.

As far as the authors are aware, *PadSteg* (Padding Steganography), presented in this paper, is the first steganographic system that utilizes what we have defined as *inter-protocol* steganography i.e. usage of relation between two or more different network protocols to enable secret communication – *PadSteg* utilizes Ethernet (IEEE 802.3), ARP, TCP and other protocols. This paper is an extension of the work introduced in [16].

Thus, classification introduced above may be further expanded to incorporate *inter-protocol* steganographic methods (Fig. 1).

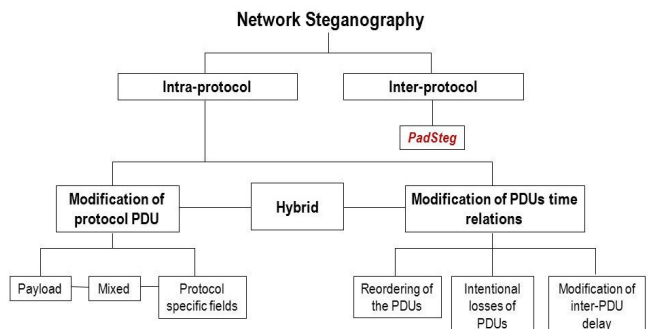

**Figure 1. Network steganography classification**

ARP (Address Resolution Protocol) [10] is a simple protocol which operates between the data link and network layers of the OSI (Open Systems Interconnection) model. In IP networks it is used mainly to determine the hardware MAC (Media Access Control) address when only a network protocol address (IP address) is known. ARP is vital for proper functioning of any switched LAN (Local Area Network) although it can raise security concerns e.g. it may be used to launch an ARP Poisoning attack.

In Ethernet, frame length is limited to a minimum of 64 octets, due to the CSMA/CD (Carrier Sense Multiple Access/Collision Detection) mechanism, and a maximum of 1500 octets. Therefore, any frames whose length is less than 64 octets have to be padded with additional data. The minimal size of an Ethernet data field is 46 octets and can be filled with data originating from any upper layer protocol, without encapsulation via the LLC (Link Layer Control), because LLC (with its 8 octets header) is very rarely utilized in 802.3 networks.

However, due to ambiguous standardization (RFC 894 and RFC 1042), implementations of padding mechanism in current NICs (Network Interface Cards) drivers vary. Moreover, some drivers handle frame padding incorrectly and fail to fill it with zeros. As a result of memory leakage, Ethernet frame padding may contain portions of kernel memory. This vulnerability is discussed in *Atstake* report and is called *Etherleak* [9]. Data inserted in padding by *Etherleak* is considered unlikely to contain any valuable information; therefore it does not pose serious threat to network security as such. However, it creates a perfect candidate for a carrier of the steganograms, thus it may be used to compromise network defenses. Utilization of padding in Ethernet frames for steganographic purposes was originally proposed by Wolf [13]. If every frame has padding set to zeros (as stated in standard), its usage will be easy to detect. With the aid of *Etherleak*, this information hiding scheme may become feasible as it will be hard to distinguish frames affected by *Etherleak* from those with steganogram.

In this paper we propose a new steganographic system *PadSteg*, which can be used in LANs and utilizes ARP and other protocols (like TCP or ICMP) together with an *Etherleak* vulnerability. We conduct a feasibility study for this information hiding system, taking into account the nature of todays' networks. We also suggest possible countermeasures against *PadSteg*.

The rest of the paper is structured as follows. Section 2 describes the *Etherleak* vulnerability and related work with regard to the application of padding for steganographic purposes. Section 3 includes a description of *PadSteg* components. Section 4 presents experimental results for real-life LAN traffic which permit for an evaluation of feasibility of the proposed solution. Section 5 discusses possible methods of detection and/or elimination of the proposed information hiding system. Finally, Section 6 concludes our work.

## II. RELATED WORK

### A. The Etherleak vulnerability

The aforementioned ambiguities within the standardization cause differences in implementation of the padding in Ethernet frames. Some systems have an implemented padding operation inside the NIC hardware (so called *auto padding*), others have it in the software device drivers or even in a separate layer 2 stack.

In the *Etherleak* report Arkin and Anderson [9] presented in details an Ethernet frame padding information leakage problem. They also listed almost 50 device drivers from Linux 2.4.18 kernel that are vulnerable.

Due to the inconsistency of padding content of short Ethernet frames (its bits should be set to zero but in many cases they are not), information hiding possibilities arise. That is why it is possible to use the padding bits as a carrier of steganograms.

Since Arkin and Anderson's report dates back to 2003, we performed an experiment in order to verify whether *Etherleak* is an issue in today's networks. The achieved results confirmed that many NICs are still vulnerable (see experimental results in Section 4).

### B. Data hiding using padding

Padding can be found at any layer of the OSI RM, but typically it is exploited for covert communications only in the data link, network and transport layers.

Wolf in [13], proposed a steganographic method which utilizes padding of 802.3 frames. Its achievable steganographic bandwidth is up to 45 bytes/frame.

Fisk et al. [7] presented padding of the IP and TCP headers in the context of active wardens. Each of these fields offers up to 31 bits/packet for steganographic communication.

Padding of IPv6 packets for information hiding was described by Lucena et al. in [8] and offers a couple of channels with a steganographic bandwidth up to 256 bytes/packet.

## III. IMPROPER ETHERNET FRAME PADDING IN REAL-LIFE NETWORKS

Real network traffic was captured to verify whether described in 2003 *Etherleak* vulnerability is still feasible in current LANs. It will also be used to evaluate the proposed in Section IV steganographic system – its steganographic bandwidth and detectability.

The experiment was conducted at the Institute of Telecommunications at Warsaw University of Technology between 15 and 19 of March 2010 (from Monday to Friday). It resulted in about 37 million packets captured, which corresponds, daily, to 7.43 million frames on average (with a standard deviation 1.2 million frames) – for details see Table 1. The traffic was captured with the aid of *Dumpcap* which is part of the *Wireshark* sniffer ver. 1.3.3 (www.wireshark.org). The sources of traffic were ordinary computer devices placed in several university laboratories and employees' ones but also peripherals, servers and network equipment. To analyze the captured traffic and calculate statistics *TShark* (which is also part of *Wireshark*) was utilized. Statistics were calculated per day, and average results are presented.

TABLE I. THE NUMBER OF CAPTURED FRAMES PER DAY

| Date | Monday | Tuesday | Wednesday | Thursday | Friday |
|---|---|---|---|---|---|
| No. of frames | 7,205,904 | 7,027,170 | 5,761,723 | 8,241,832 | 8,945,403 |

The captured traffic classification by upper layer protocol is presented in Fig. 2. Three quarters of the traffic was HTTP. Together with SSH, UDP and SSL protocols it sums up to about 93% of the traffic.

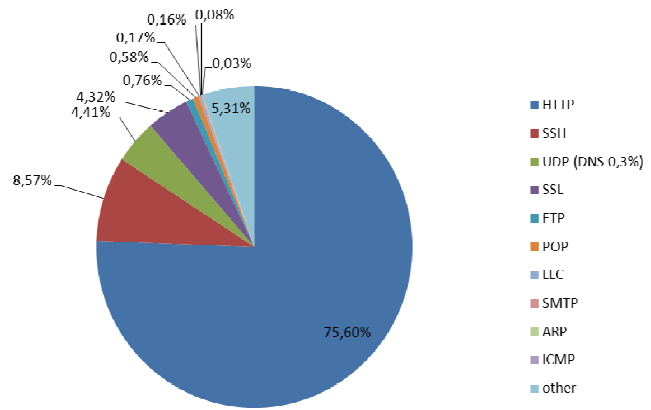

**Figure 2. Captured traffic characteristics**

Almost 22% (with a standard deviation of 7.7%) of all daily traffic had padding bits added (~8 million frames). It is obvious that not all of the frames were affected since padding is added only to small-sized packets.

Table 2 shows for which network protocols frames were mostly improperly padded.

TABLE II. UPPER LAYER PROTCOLS AFFECTED WITH ETHERNET FRAME IMPROPER PADDING IN EXPERIMENTAL DATA AND EXEMPLARY PID ASSIGMENT

| Affected protocol | TCP | ARP | ICMP | UDP | Others |
|---|---|---|---|---|---|
| [%] | 92.82 | 4.17 | 2.31 | 0.54 | 0.16 |
| PID | 1 | 2 | 3 | 4 | - |

However, it is important to note, that almost 22% of the padded frames experienced improper padding (~1.8 million frames). These frames were generated by about 15% of hosts in the inspected network (their NICs were produced among others by some US leading vendors). We considered Ethernet frame padding improper if the padding bits were not set to zeros.

TCP segments with an ACK flag set (which have no payload) result in frames that have to be padded, thus, it is no surprise that ~93% of improperly padded traffic is TCP. Nearly all of this traffic consists of ACK segments. Other frames that had improper padding were caused by ARP and ICMP messages – *Echo Request* and *Echo Reply* (~6.5%). It is also worth noting that there is also padding potential in UDP datagrams as UDP-based applications often generate small-sized frames (e.g. voice packets in IP telephony). However, padding was only present in 0.5% of all padded frames.

For *PadSteg* ARP protocol plays important role (see Section IV for details), thus our aim was also to find out ARP statistics i.e. what are the most frequently used ARP messages, what is their distribution and how many of them have improper padding. The results are presented in Fig. 3.

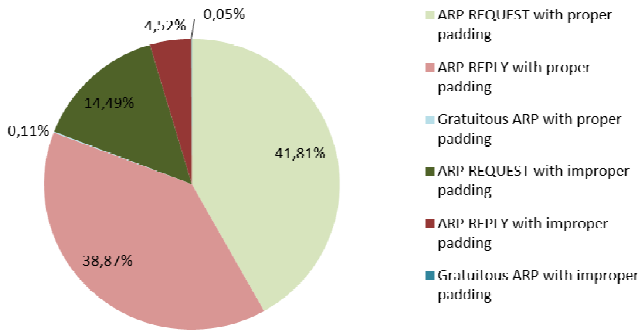

**Figure 3. Captured ARP characteristics**

Not surprisingly, the most frequently sent ARP messages were ARP Request (~56.3%) and Reply (~43.4%), while Gratuitous ARP messages are in minority (~0.2%). Out of all ARP messages almost 20% had improper padding.

## IV. COMPONENTS OF THE PROPOSED STEGANOGRAPHIC SYSTEM

*PadSteg* enables secret communication in a hidden group in a LAN environment. In such group, each host willing to exchange steganograms should be able to locate and identify other hidden hosts. To provide this functionality certain mechanisms must be specified. In our proposal, ARP protocol, together with improper Ethernet frame padding are used to provide localization and identification of the members of a hidden group. To exchange steganograms improper Ethernet frame padding is utilized in frames that in upper layer use TCP, ARP or ICMP (or other network protocols that cause Ethernet frames to be padded). These protocols will be called carrier-protocols as they enable transfer of steganograms throughout the network.

Moreover, while the secret communication takes place, hidden nodes can switch between carrier-protocols to minimize the risk of disclosure. We called such mechanism *carrier-protocol hopping* and it will be described in details later.

In this section we first describe ARP protocol, and then we focus on proposed steganographic system operations.

### A. Overview of ARP Protocol

ARP returns the layer 2 (data link) address for a given layer 3 address (network layer). This functionality is realized with two ARP messages: Request and Reply. The ARP header is presented in Fig. 4.

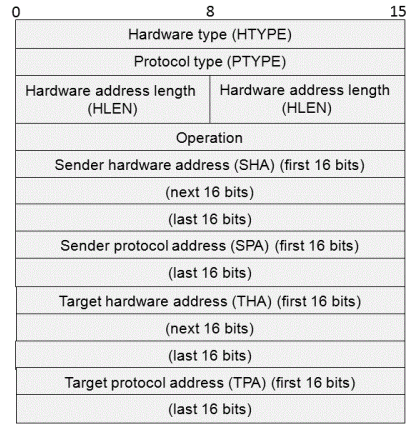

**Figure 4. ARP header format**

ARP header fields have the following functions:
- HTYPE (Hardware Type) – type of data link protocol used by sender (1 is inserted if it is Ethernet).
- PTYPE (Protocol Type) – type of network protocol in network layer (0800h is inserted if IP is used).
- HLEN (Hardware Length) – length of hardware address fields: SHA, THA (in bytes).

- PLEN (Protocol Length) – length of protocol address fields: SPA, THA (in bytes).
- OPER (Operation) – defines, whether the frame is an ARP REQUEST (1) or REPLY (2) message.
- SHA (Sender Hardware Address) – sender data link layer address (MAC address for Ethernet).
- SPA (Sender Protocol Address) – sender network layer address.
- THA (Target Hardware Address) – data link layer address of the target. This field contains zeros whenever a REQUEST ARP message is sent.
- TPA (Target Protocol Address) – network layer address of the target. This field contains zeros if REQUEST ARP message is sent.

An example of ARP communication with Request/Reply exchange, captured with the *Wireshark* sniffer (www.wireshark.org), is presented in Fig. 5. First, ARP Request is issued (1), which is used by the host with IP address 10.7.6.29 to ask other stations (by means of broadcast): 'Who has IP 10.7.56.47?'. In order to send a frame intended for everyone in a broadcast domain, Ethernet header destination address must be set to FF:FF:FF:FF:FF:FF (2). Next, host with IP address 10.7.56.47 replies directly to 10.7.6.29 using unicast ARP Reply (3) with its MAC address.

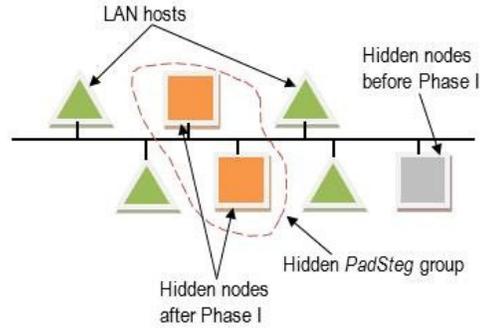

**Figure 5. ARP exchange captured with *Wireshark***

Basing on the proposed description of ARP protocol, it can be concluded that ARP header is rather of fixed content and presents little possibilities for information hiding. One opportunity is to modulate address fields like it was proposed in [11] or [8]. However, this solution provides limited steganographic bandwidth if certain level of undetectability is to be achieved. Moreover, it may result in improper IP and MAC address advertisements which may make this method more prone to detection.

Thus, in the proposed steganographic system *PadSteg*, we utilize ARP Request messages, broadcasted throughout LAN, to make other members of the hidden group become aware of the presence of a new member.

### B. Steganographic system operation

*PadSteg* is designed for LANs only because it utilizes improper Ethernet frame padding in Ethernet. It allows members of the hidden groups to secretly exchange data (Fig. 6).

**Figure 6. *PadSteg* hidden group**

Every member from the hidden group is obligated to fill each short Ethernet frame it sends with non-zero padding to make detection harder – such node must mimic *Etherleak* vulnerability. *PadSteg* also uses protocols like ARP, TCP or ICMP to control hidden group and to transfer steganograms.

*PadSteg* operation can be split into two phases:
- Phase I: Advertisement of the hidden node and a carrier-protocol.
- Phase II: Hidden data exchange with optional carrier-protocol change.

*Phase I*

This phase is based on the exchange of ARP Request messages with improper Ethernet frame padding (Fig. 7).

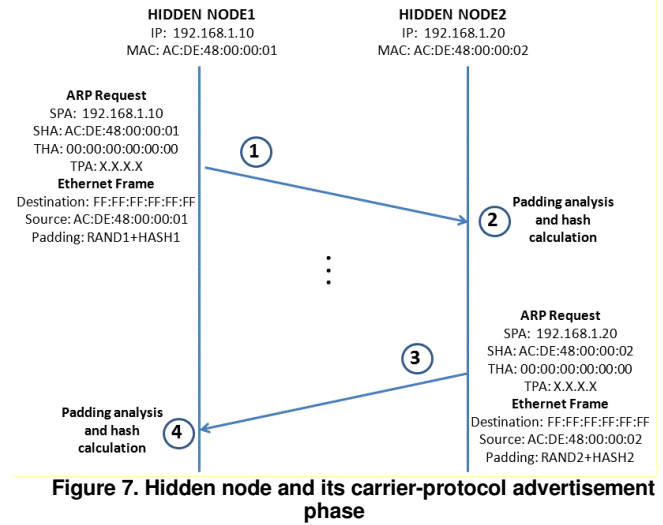

**Figure 7. Hidden node and its carrier-protocol advertisement phase**

Hidden node that wants to advertise itself to others in the group, broadcasts an ARP Request message (1) and inserts *advertising sequence* into the padding bits. It consists of: a random number *RD* (different from 0), and hash $R_H$ which is calculated based on *RD*, carrier-protocol identifier *PID* and source MAC address (see eq. 4-1). Incorporating *RD* ensures that frame padding will be random. *PID* is an identifier of the upper layer carrier-protocol for the steganograms transfer and may have been assigned exemplary values like in Table II. *PID* is used to advertise hidden node preference for the secret data transfer and may be used during steganograms exchange by carrier-protocol hopping mechanism.

An example of the padding bits format (which for ARP is 144 bits long), assuming usage of MD5 hash function, is presented in Fig. 8.

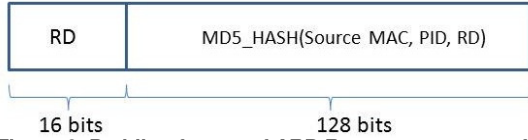

**Figure 8. Padding format of ARP Request messages for the activation phase**

All the hidden nodes are obligated to analyze the padding of all received ARP Requests. If an ARP Request is received with padding that is not all zeros, it is analyzed by extracting the random number and calculating corresponding hashes (2) as follows

$$R_H(PID) = H(PID \| RD \| SR\_MAC) \quad (4-1)$$

For each extracted hash, receiver computes hashes with different *PID*. The order of the *PID* values for hashes calculation should correspond to traffic characteristics i.e. more likely carrier-protocols should be checked first. For example, based on *PID* values in Table II, $R_H(1)$ will be computed first, then $R_H(2)$ etc. because padding will more likely occur for TCP protocol than ARP and others. Such approach will limit unnecessary hashes calculation. Finally, if the received and calculated hashes are the same it means that a new hidden node is available for steganographic exchange and the carrier-protocol for this node is established. It means that if any hidden node receives frames from this new hidden node, only these corresponding to extracted *PID* value carry steganogram and will be analyzed.

Each hidden node stores a list of nodes from which it has received advertisements with their advertised carrier-protocol. Every hidden node should also reissue ARP Requests at certain time intervals to inform other hidden nodes about its existence. To limit the chance of detection, sending of ARP Requests may not happen too often (3, 4). In ARP, if an entry in host ARP cache is not refreshed within 1 to 20 minutes (implementation dependent) it expires and is removed. Thus, hidden nodes should mimic such behavior to imitate the sending of ARP Requests caused by ARP cache expiration.

Adaptation of ARP messages for identification of new hidden nodes has two advantages:
- The broadcast messages will be received by all hosts in LAN.
- The ARP traffic totals to about 0.1% of all traffic (see next Section for details), so this choice is also beneficial from the performance perspective. Each hidden node does not have to analyze all of the received traffic but only ARP Requests.

*Phase II*

After the identification of a new hidden node and its carrier-protocol, other hidden nodes analyze each short Ethernet frame's padding sent from that MAC address that in upper layers has chosen carrier-protocol. The received frames' padding contains steganogram bits.

The bidirectional transmission is performed as presented in Fig. 9. Two hidden nodes make e.g. an overt TCP connection – they transfer a file (1). During the connection TCP ACK segments are issued with improper Ethernet frame padding (2 and 4). Received TCP segments are analyzed for improper Ethernet padding presence and secret data is extracted (3 and 5). For third party observer such communication looks like usual data transfer.

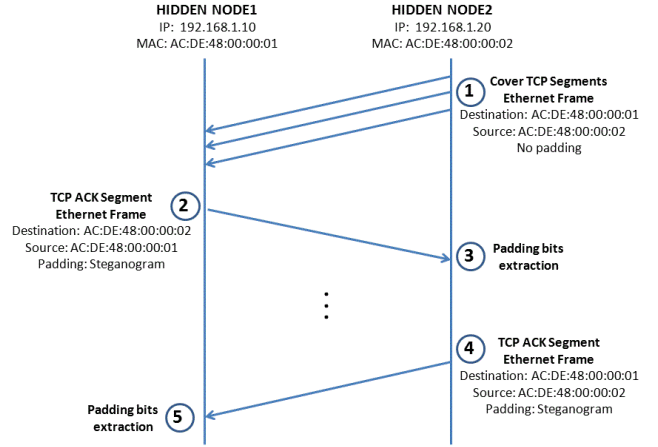

**Figure 9. Hidden group steganograms exchange phase**

During the exchange of steganograms or between two consecutive connections between two hidden nodes changing of carrier-protocol may occur. Hidden nodes may achieve this with use of *carrier-protocol hopping* mechanism. Let assume that there are two hidden nodes HN1 and HN2 and they want to change their carrier-protocols. To achieve it they do as follows (see Fig. 10):
- When HN1 wants to change its carrier-protocol it issues ARP Request which contains different from previous *PID* included in the hash inserted into the padding of this frame (see Fig. 8). ARP Request has TPA field set to IP address of the HN2 (1).
- After receiving ARP Request HN2 updates its list of hidden nodes and their carrier-protocols based on calculated hash analysis and PID (2). Then HN2 issues ARP Reply directly to HN1, which in padding contains its carrier-protocol preference (3).
- When HN1 receives ARP Reply it updates its list of hidden nodes and their carrier-protocols and is ready to use different carrier-protocol for HN2 i.e. it will analyze padding from all the short frames that in upper layers has chosen carrier-protocol (4).

Note that steganogram exchange does not necessarily must be symmetrical i.e. hidden nodes do not have to use the same carrier-protocols which performing hidden data transfer.

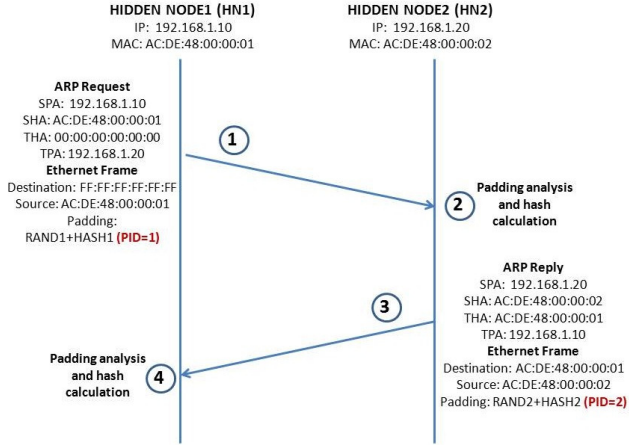

**Figure 10. Carrier-protocol hopping mechanism example**

## V. PADSTEG EVALUATION

### A. Padding content analysis

Table III presents hexadecimal values of frame padding, written in regular expression standard. Depending on day of observation padding contained different values, therefore we cannot state which value occurred most or least often. However, values bolded did not change in consecutive days. Some values were constant and other completely random. Therefore, we can make an assumption that padding content pattern changes with reboot of the device. Results confirm that memory leakage value in padding show some patterns that are very difficult to predict. That is why, we suggest that the proposed system should sacrifice few bits of the padding to generate some pattern in every message in order to increase undetectability.

TABLE III. FRAME PADDING CONTENT VARIETY (HEXADECIMAL VALUES)

| Padding Length | 6B | 18B |
|---|---|---|
| Regex | **00{2}[0-F]{4}** | **80fca7a0[0-F]{14}** |
| | **80[0-F]{5}** | **a96f[0-F]{16}** |
| | c0[0-F]{5} | 00{14} [0-F]{4} |
| | 20{6} | [0-F]+00{3}[0-F]* |
| | 474554202f[0-F]{1} | 80fca7a0ffffffffffff[0-F]{8} |
| | 0101050a74b6 | 80fca7a080fe88e0ffffffff0012179cfd53 |
| | [0-F]{6} (random) | [0-F]{18} (random) |

### B. Steganographic bandwidth estimation

Let us try to estimate *PadSteg* steganographic bandwidth for a single hidden node transmitting in a hidden group.

Because, currently, there are no tools for steganography detection, in real-life networks, every member of a hidden group can exchange almost unlimited number of steganograms and remain undiscovered. However, if the network traffic is consequently monitored, a naive use of *PadSteg* – that is: excessive generation of Ethernet frames with improper padding may be easily detected.

This leads to conclusion that it is important to evaluate what is the realistic steganographic bandwidth under the assumption that the secret data exchange will not differ from other hosts' traffic burdened with the *Etherleak* vulnerability. To achieve this goal steganographic user's network activity must mimic behavior of other users in terms of sending Ethernet frames with improper padding.

We calculated the steganographic bandwidth of the proposed system based on the average, daily number of TCP, ARP, ICMP, UDP messages with improper Ethernet padding per susceptible host (see Table IV).

Because each TCP and ICMP messages padding is 6 bytes long, ARP message padding 18 bytes, the average steganographic bandwidth is about 32 bit/s (with a daily standard deviation of about 14 bit/s). Therefore, if the hidden node generates Ethernet frames with improper padding that fall within the average range, for the inspected LAN network, steganographic communication may remain undetected.

TABLE IV. THE NUMBER OF FRAMES WITH IMPROPER PADDING PER HOST

| Prot. | Monday | Tuesday | Wednesday | Thursday | Friday |
|---|---|---|---|---|---|
| TCP | 25,379 | 53,469 | 31,014 | 79,981 | 52,940 |
| ARP | 1,036 | 250 | 2,116 | 2,828 | 1,825 |
| ICMP | 618 | 1,330 | 1,154 | 1,660 | 9 |
| UDP | 31 | 117 | 65 | 1,773 | 77 |

TABLE V. ESTIMATED STEGANOGRAPHIC BANDWIDTH

| [bit/s] | TCP | ARP | ICMP | Sum |
|---|---|---|---|---|
| Average steg. bandwidth | 26.98 | 3.43 | 1.90 | 32.31 |
| Standard deviation | 12.03 | 1.15 | 0.66 | 13.84 |
| Confidence Interval (95%) | 5.41 | 0.52 | 0.30 | 6.23 |

### C. PadSteg prototype

*PadSteg* prototype – *StegTalk* – was implemented in C/C++ programming language with use of WinPcap 4.1.1 library (www.winpcap.org) for Windows XP OS. *StegTalk* is limited in functioning to ARP protocol only, so the PID value (see Fig. 8) is constant and equal 2. Application allows sending and receiving content from *.txt files between program instances running on different hosts.

*StegTalk* behavior is not deterministic in time. Messages containing steganograms are sent every ~60 seconds (depending on initial command line arguments) and initialization messages every 180 seconds, imitating host with Windows XP OS behavior. The ~60 seconds interval was estimated in the following way. Based on experimental results presented in Table V maximum steganographic throughput that sustains high undetectability level, using ARP protocol is ~4 bit/s. It means that a single ARP message is issued every ~45 seconds. However, because initialization ARP messages are sent every 180 seconds, therefore, messages containing actual data should be sent every ~60 seconds.

Exemplary *StegTalk* output and functioning is presented in Fig. 10. Hidden host received ARP message and discovered new hidden node (1). Then host sent its own advertisement ARP message with steganographic capabilities (2). Every ARP message that hash was not successfully recognized is ignored (3). Each ARP message which is received from known hidden node is verified and hidden data is extracted ("topsecretmessage") (4).

Figure 10. StegTalk application functioning

*StegTalk* tests were conducted on two virtual PC's with use of VMware Server 2.0 (www.vmware.com). Fixed-size text was sent from one host to another three times for each application mode (maximizing undetectability --*slow* or throughput --*fast,* see Fig. 11), in order to measure the time needed to receive the full text. Measured goodput (application level throughput) was approx. 2.3 bit/s and depending on program initial command line arguments it varied between 1.7 bit/s and 2.5 bit/s (standard deviation approx. 0.2 bit/s).

Figure 11. StegTalk application arguments

Having tested *StegTalk* behavior, in order to estimate application undetectability, sample host's network traffic had to be profiled – Fig. 12. Generally, application generates significantly fewer messages than the host during each 24h period. It is worth noting that the total amount of ARP messages will be a sum of those generated by host and *StegTalk*. Editing Windows OS registry keys may decrease the amount of ARP messages send by host and would increase *StegTalk* undetectability.

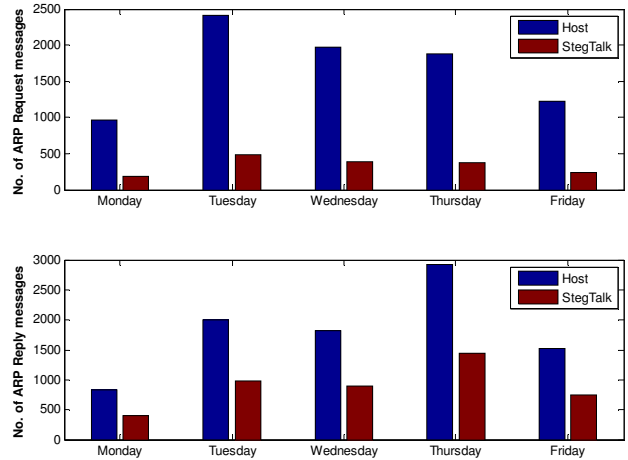

Figure 12. No. of ARP messages generated each day by an exemplary host and *StegTalk* application

VI. POSSIBLE COUNTERMEASURES

Our proposal of the new steganographic system, *PadSteg*, proves that such phenomenon like *inter-protocol steganography* is possible and may pose a threat to network security.

In today's LANs, with security measures they provide, *PadSteg* will be hard to detect. The main reason for this is that current IDS/IPS (Intrusion Detection/Prevention System) systems are rarely used to analyze all traffic generated in a LAN as this would be hard to achieve from the performance point of view. Moreover, usually IDSs/IPSs operate on signatures, therefore they require continuous signatures updates of the previously unknown steganographic methods, especially, if the information hiding process is distributed over more than one network protocol (as it is in *PadSteg*).

Thus, the best steps we can take to alleviate *PadSteg* in LANs are to:
- Ensure that there are no NICs with *Etherleak* vulnerability in the LAN.
- Enhance IDS/IPS rules to include *PadSteg* and deploy them in LANs.
- Improve access devices (e.g. switches) by adding active warden functionality [7] i.e. ability to modify (set to zeros) Ethernet frame padding if an improper one is encountered.

Implementation of the specified countermeasures greatly minimizes the risk of successful *PadSteg* utilization.

## VII. CONCLUSIONS

In this paper we presented new steganographic system - *PadSteg* – which is the first information hiding solution based on *inter-protocol steganography*.

It may be deployed in LANs and it utilizes two protocols to enable secret data exchange: Ethernet and ARP/TCP. A steganogram is inserted into Ethernet frame padding but one must always "look" at the other layer protocol (ARP or TCP) to determine whether it contains secret data or not. Based on the results of conducted experiment the average steganographic bandwidth of *PadSteg* was roughly estimated to be 32 bit/s. It is a quite significant number considering other known steganographic methods.

In order to minimize the potential threat of *inter-protocol steganography* to public security identification of such methods is important. Equally crucial is the development of effective countermeasures. This requires an in-depth understanding of the functionality of network protocols and the ways in which they can be used for steganography.

However, considering the complexity of network protocols being currently used, there is not much hope that a universal and effective steganalysis method can be developed. Thus, after each new steganographic method is identified, security systems must be adapted to the new, potential threat.

As a future work larger volumes of traffic from different LANs should be analyzed in order to pinpoint more accurately *PadSteg* feasibility and calculate its steganographic bandwidth.

ACKNOWLEDGMENT

This work was partially supported by the Polish Ministry of Science and Higher Education under Grant: N517 071637.

REFERENCES

[1] Rowland C., Covert Channels in the TCP/IP Protocol Suite, First Monday, Peer Reviewed Journal on the Internet, July 1997
[2] Zander S., Armitage G., Branch P., A Survey of Covert Channels and Countermeasures in Computer Network Protocols, IEEE Communications Surveys & Tutorials, 3rd Quarter 2007, Volume: 9, Issue: 3, pp. 44-57, ISSN: 1553-877X
[3] Petitcolas F., Anderson R., Kuhn M., Information Hiding – A Survey: IEEE Special Issue on Protection of Multimedia Content, July 1999
[4] Murdoch S.J., Lewis S., Embedding Covert Channels into TCP/IP, Information Hiding (2005), pp. 247-26
[5] Ahsan, K. and Kundur, D.: Practical Data Hiding in TCP/IP, Proc. ACM Wksp. Multimedia Security, December 2002.
[6] Kundur D. and Ahsan K.: Practical Internet Steganography: Data Hiding in IP, Proc. Texas Wksp. Security of Information Systems, April 2003.
[7] Fisk, G., Fisk, M., Papadopoulos, C., Neil, J.: Eliminating Steganography in Internet Traffic with Active Wardens, In Proc: 5th International Workshop on Information Hiding, Lecture Notes in Computer Science: 2578, 2002, str. 18–35
[8] Lucena N. B., Lewandowski G., Chapin S. J., Covert Channels in IPv6, Proc. Privacy Enhancing Technologies (PET), May 2005, pp. 147–66.
[9] Arkin O., Anderson J., Ethernet frame padding information leakage, Atstake report, 2003
http://packetstorm.codar.com.br/advisories/atstake/atstake_etherleak_report.pdf
[10] Plummer D. C., An Ethernet Address Resolution Protocol, RFC 826, November 1982
[11] Girling C. G., Covert Channels in LAN's, IEEE Trans. Software Engineering, vol. SE-13, no. 2, Feb. 1987, pp. 292–96.
[12] Handel T., Sandford M.. Hiding Data in the OSI Network Model. In Proceedings of the First International Workshop on Information Hiding, pages 23-38, 1996.
[13] Wolf M, "Covert Channels in LAN Protocols," Proc. Wksp. Local Area Network Security (LANSEC), 1989, pp. 91–101.
[14] Mazurczyk W, Szczypiorski K (2008) Steganography of VoIP Streams, In: R. Meersman and Z. Tari (Eds.): OTM 2008, Part II - Lecture Notes in Computer Science (LNCS) 5332, Springer-Verlag Berlin Heidelberg, Proc. of The 3rd International Symposium on Information Security (IS'08), Monterrey, Mexico, November 2008, pp. 1001-1018
[15] Mazurczyk W., Smolarczyk M., Szczypiorski K.: Retransmission steganography and its detection, Soft Computing, ISSN: 1432-7643 (print version), ISSN: 1433-7479 (electronic version), Journal no. 500 Springer, November 2009
[16] B. Jankowski, W. Mazurczyk, K. Szczypiorski, Information Hiding Using Improper Frame Padding, Submitted to 14th International Telecommunications Network Strategy and Planning Symposium (Networks 2010), 27-30.09.2010, Warsaw, Poland

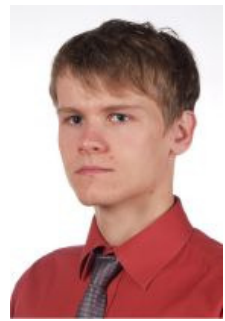
Bartosz Jankowski studies telecommunication at Warsaw University of Technology (WUT, Poland) since 2007. His main areas of interest are network security, information hiding techniques and recently project management. Member of the Network Security Group at WUT (secgroup.pl) and coauthor of first inter-protocol steganographic system *PadSteg*. He is regarded as goal-oriented person with a strong drive to learn. He is a co-author of 3 publications and 1 invited talk.

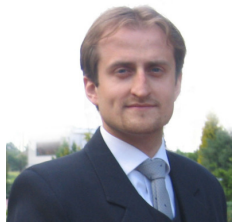
Wojciech Mazurczyk holds an M.Sc. (2004) and a Ph.D. (2009) in telecommunications from the Faculty of Electronics and Information Technology, Warsaw University of Technology (WUT, Poland) and is now an Assistant Professor at WUT and the author of over 50 scientific papers and over 25 invited talks on information security and telecommunications. His main research interests are information hiding techniques, network security and multimedia services. He is also a research co-leader of the Network Security Group at WUT (secgroup.pl). Personal website: http://mazurczyk.com.

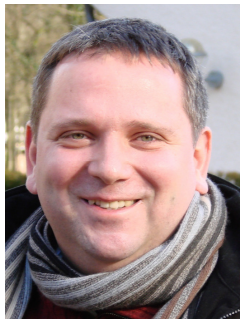

Krzysztof Szczypiorski holds an M.Sc. (1997) and a Ph.D. (2007) in telecommunications both with honours from the Faculty of Electronics and Information Technology, Warsaw University of Technology (WUT), and is an Assistant Professor at WUT. He is the founder and head of the International Telecommunication Union Internet Training Centre (ITU-ITC), established in 2003. He is also a research leader of the Network Security Group at WUT (secgroup.pl). His research interests include network security, steganography and wireless networks. He is the author or co-author of over 110 publications including 65 papers, two patent applications, and 35 invited talks.